\documentclass{article}

\usepackage[english]{babel}
\usepackage{float}
\usepackage[letterpaper,top=2cm,bottom=2cm,left=3cm,right=3cm,marginparwidth=1.75cm]{geometry}

\usepackage{amsmath}
\usepackage{graphicx}
\usepackage[colorlinks=true, allcolors=blue]{hyperref}
\title{Thomas Attractor}
\author{
  Idan Sorin 
  Michael Tulchinsky \\
  Faculty of Mathematics 
  Technion Israel Institute of Technology 
  Haifa 32000, Israel
}

\begin{document}
\maketitle
\begin{abstract}
 
An analytical and numerical analysis presented in the paper of a simple symmetric dynamical system that exhibits rich behavior.
The system is cyclical in the Cartesian coordinates. Moreover, it has only one control parameter which governs the behavior of the system. 
Depending on the value of the control parameter, the system can exhibit vastly different behavior patterns such as a stable regime, infinite limit cycles, infinite amount of bifurcations, a series growing to infinity of fixed points, and chaos containing multiple attractors. \\ In the extreme case where the control parameter goes to zero, thus there is no dissipation term, the system exhibits an infinite number of fixed points and behaves like a Brownian motion. \\
We would like to thank Dr. Tali Pinsky from the Technion for her guidance and help in writing this article.
\end{abstract}

\section{Introduction}

Thomas system is described by the equations:

\begin{equation}
\begin{aligned}
\dot{x} &= \sin(y) - b x \\
\dot{y} &= \sin(z) - b y \\
\dot{z} &= \sin(x) - b z
\end{aligned}
\tag{1}
\end{equation}
where \(b > 0\) is the control parameter of the system. \\
The system has unique properties, which makes it very different from other famous systems such as Lorenz [4] and Rossler [3]. \\
First, it contains only one bifurcation parameter, that governs all the dynamics. Moreover, the derivatives contain \( \sin \) and not polynomials, which ensures rich dynamics containing the possibility of a multitude of fixed points and the coexistence of multiple attractors for different values of the control parameter. \\

As was mentioned in Sprott [2], the physical interpretation of the system is of a particle that moves under a cyclically symmetric potential while also exhibiting a damping force. In other words, we can expect the particle to move in counterintuitive ways in a maze-like structure while shedding energy in the process. As for the systems' appearance in other disciplines, the system is a simplification of auto-catalytic models which are prevalent in evolution [8], chemical reactions [7], and ecology [15]. \\

Throughout the years, attempts were made to analyze different aspects of the system at hand and this paper borrows and expands on such attempts by Sprott [2] and R.Thomas [1]. In the paper, we only briefly mention the dynamics of the no-damping scenario as it was extensively worked on in previous papers. \\
We start off with analytical analysis such as finding the fixed points, and analyzing their stability including finding a Lyapunov function for the origin. \\

Afterward, we analyze a sample from the bifurcations of the system and move to the numerical analysis, finding the chaos regime of the system by a bifurcation diagram, finding the Lyapunov exponents, Kaplan-Yorke dimension of the attractors and a Poincare map. We also looked at the special case \( b=0 \) and analyzed it analytically and numerically. \\
We find that while the control parameter varies from one to zero, the system exhibits infinite "Double saddle-nodes" bifurcations as well as infinite Hopf bifurcations, generating infinite stable limit cycles inside the chaotic regime. We also found a systematic way for finding all those bifurcations. An interesting behavior can be seen qualitatively where for a critical value of the control parameter the attractors go from being separable into a mixing state which gets more violent as the control parameter goes to zero.  
   
\section{Analytical Analysis }

Looking at system (1), one can notice that there is a symmetry of cyclic changing variables (x,y,z) i.e, one can change the names of the parameters in a cyclic way and get the same equations. One can conclude that the solutions of the system (fixed points, limit cycles, and attractors) should also maintain this symmetry, so all the fixed points must be from the form (x*,x*,x*). This gives us a much simpler way to find them- except of solving 3-D non-linear system of equations, we can just look at one of the equations, for example, the first, change the name of the variables to the same name, and solve. Thus, all the fixed points can be found from the equation: \[bx=sin(x) \ (2)\]  \\
Note that we also have a reflection symmetry, which means that every phenomenon that happens on the right side of the axis, will happen on the left side of the axis. \\
Notice that as we decrease b, more fixed points can occur and when \(b\rightarrow 0\) the number of the fixed points is not bounded, so we get infinite bifurcations from a single parameter. \\
The Jacobian matrix of the system is \[
J = \begin{bmatrix}
-b & \cos(y) & 0 \\
0 & -b & \cos(z) \\
\cos(x) & 0 & -b \\
\end{bmatrix}
\]
and in the fixed points (x*,x*,x*) it is getting the form 

\begin{equation}
J = \begin{bmatrix}
-b & c & 0 \\
0 & -b & c \\
c & 0 & -b \\
\end{bmatrix}
\tag{3}
\end{equation}

when \(c=cos(x*)\). \\
Note that in our case, the Jacobian matrix is a circulant one, so we can find its eigenvalues directly: 
\begin{equation}
\lambda_0 = c - b, \quad \lambda_{1,2} = -(b + \frac{c}{2}) \pm \ \frac{1}{2} \sqrt{3}c i
\tag{4}
\end{equation}

\subsection{The case \texorpdfstring{\(b > 1\)}{b > 1}}

When \(b>1\), equation (2) has only one solution, which is the origin (0,0,0). \\
since \(c=cos(0)=1\), we see that the eigenvalues from equation (4) all have a negative real part, so the origin is a stable spiral.
We would like to prove global stability, so we consider the function:
\[
V(x, y, z) = \frac{1}{2}(x^2 + y^2 + z^2)
\]
Thus, \[
\dot{V}(x, y, z) = -b(x^2+y^2+z^2)+xsin(y)+ysin(z)+zsin(x)
\]
Of course, V is greater than zero and equals to zero only in the origin. We want to prove that its derivative is always negative. We can see that if all the variables are greater than 1 in their absolute value, the derivative is always negative since every quadratic term is greater than every \(sin\) term in absolute value. So, it is enough to look at the case when at list one of the variables is bounded between -1 and 1, for example, z. Now we look at the surface   
  \[
(x^2 + y^2 + z^2)= xsin(y)+ysin(z)+zsin(x)
\]  
when z is considered as a parameter between -1 and 1. Numerically, for values of z between -1 and 1 the solution is only the origin.\\
Since \[-b(x^2+y^2+z^2)< -(x^2+y^2+z^2)\] for every (x,y,z), not the origin, we conclude that
\[
\dot{V}(x, y, z) \leq 0
\]
and equals zero only at the origin. Therefore, our function is a Lyapunov function, and the origin is globally stable.

\subsection{ \texorpdfstring{\(b < 1\)}{b < 1}}

at b=1, equation (2) has two more solutions, which can be approximated by Taylor series to \(sin\): \[ bx\approx x+ \frac{x^3}{6}+...\] so the solutions are \[x\approx \pm \sqrt{6(1-b)}\]
We now check the stability of the new points: for the origin, c=1 so the complex eigenvalues have always negative real part. The real eigenvalue becomes positive since b is less than 1
so the origin loses its stability.\\
For the stability of the two new fixed points, we will look at the real values of their eigenvalues. \(\lambda_0\ <0 \) iff \[cos(x*)<b= \frac{sin(x*)}{x*}\] and this relation holds for every x* close to zero. So \(\lambda_0 <0 \) when b becomes less than 1.
For \(\lambda_{1,2}\) we know that their Real is negative iff \(-(b+\frac{c}{2})<0\) iff \[-(\frac{sin(x*)}{x*}-\frac{cos(x*)}{2})<0\] and it also holds for x* close to zero, so the new two fixed points are stable. It follows that we got a pitchfork bifurcation. More and more new fixed points will occur as b decreases, because of the solutions of equation (2). The next time that new points will appear is when \(b\approx 0.13\). \\

If we analyze the fixed points as \( b \) decreases we get that after the pitchfork bifurcation, the two new fixed points lose their stability with no formation of new points. When decreasing \( b \) even more, we found that the new 4 fixed points, two couples of two fixed points symmetrically to the \( x \) axis are creating, which means that a double saddle-node bifurcation occurred. The two outside points are stable, while the two inner points are unstable. If continuing decreasing \( b \) we can see numerically that this process is periodic—the two outside points will lose their stability for some smaller \( b \) with no new fixed points formatting, and when decreasing \( b \) even more another new 4 fixed points are created, and so on. One can explain this behavior from the structure of the function \(\frac{\sin(x)}{x}\). \\ It has an infinite amount of maximum points, with a positive but decreasing to 0 \( y \) value. So, when decreasing \( b \), every time it passes the value of a new maximum point, there are another two solutions for equation (2), two values of \( x \) in the neighborhood of the maximum point. Since the function is symmetric, we get another two negative solutions so we get 4 solutions in total.

\begin{figure}[htbp]  
    \centering  
    \includegraphics[width=0.5\linewidth]{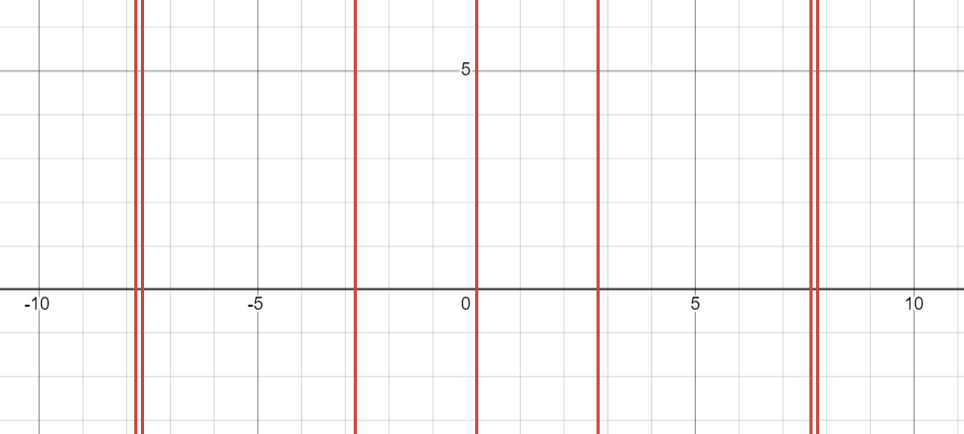}  
    \caption{The new fixed points near the bifurcation for \( b = 0.128 \)}
    \label{fig:bifurcation}
\end{figure}

\begin{figure}[htbp]  
    \centering
    \includegraphics[width=0.5\linewidth]{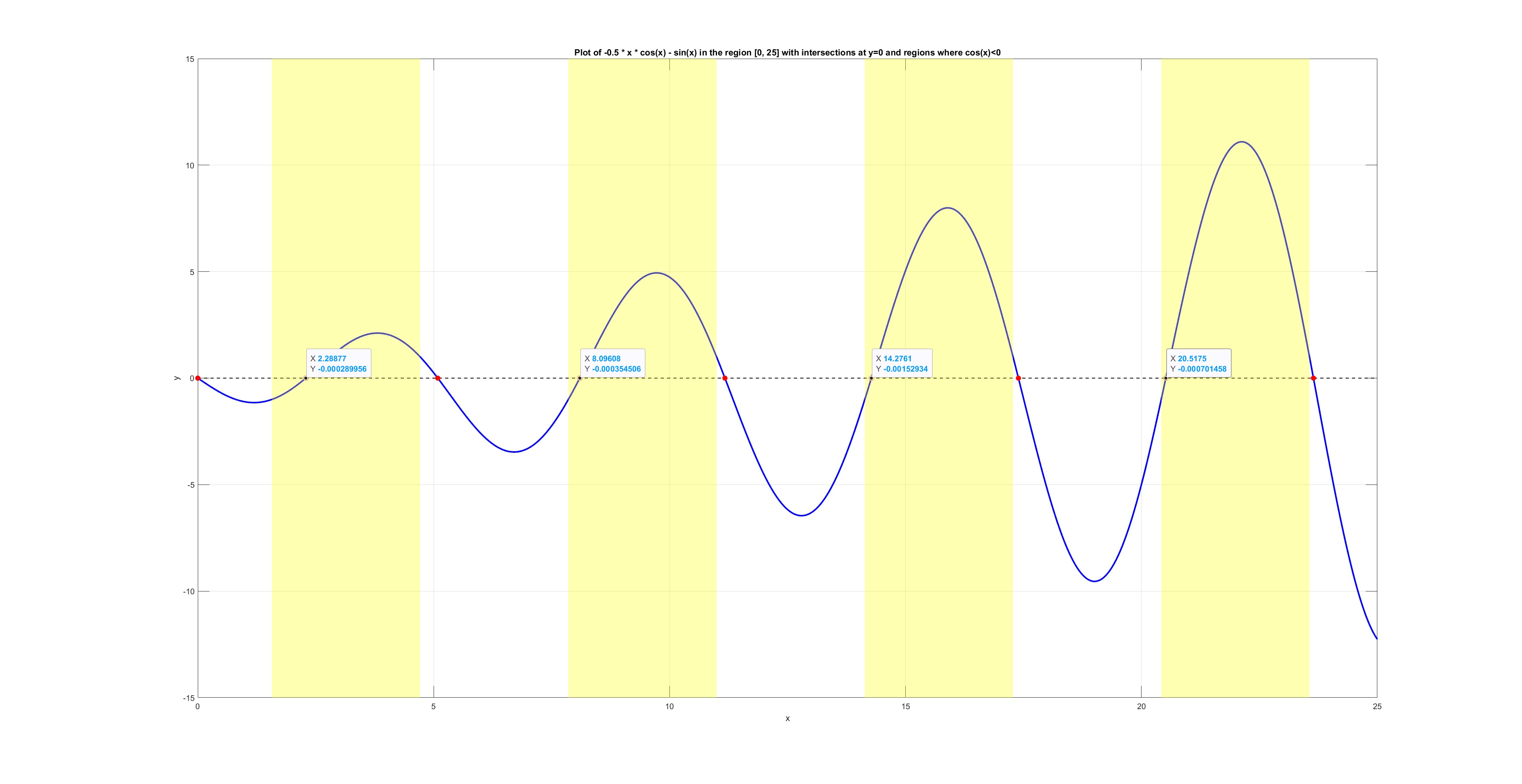}  
    \caption{Points where Hopf bifurcation occurs, \( x^* = 2.28, 8.09, 14.27, 20.51 \)}
    \label{fig:enter-label}
\end{figure}

Another bifurcation that we can get is when the real value of the complex eigenvalues becomes zero. This is a Hopf bifurcation when a limit cycle is born. It happens when \[ b = -\frac{c}{2} \] so from (2) and (3) it's equivalent to \[ -0.5x \cos(x) = \sin(x) \] this equation has infinite solutions, but we are looking for the solutions such that \( b \) is positive, i.e. \[ \cos(x^*) < 0 \] and we can see that there are infinite solutions (Figure \ref{fig:enter-label}) which means that there are infinite Hopf bifurcations and periodic windows. Also, \[ \lambda_0 = -3b < 0 \] so they are stable limit cycles. Two examples of such bifurcations are at \( b = 0.329 \) and \( b = 0.1198 \). We can conclude that even in a chaotic regime we will always have some periodic windows, so the system is not chaotic everywhere. This behavior is more typical to discrete dynamical systems than to continuous ones [16].

\section{The chaos regime}

We would like to find the regime when chaos arises. For that purpose, we first calculate the divergent:
\[
\nabla \cdot \textbf{v} = \frac{\partial \dot{x}}{\partial x} + \frac{\partial \dot{y}}{\partial y} + \frac{\partial \dot{z}}{\partial z} = -b - b - b = -3b < 0
\]
So the volume in the phase space goes to zero as t goes to infinity.
For the chaos regime, we had to use more numerical tools, such as calculating the Lyapunov exponents, the Kaplan Yorke dimension of the system, a bifurcation diagram, and a Poincare section. These tools can also ensure that our analytical results are true. We also plot some of the attractors and limit cycles (see Figure 3).

\subsection{Lyapunov exp and Kaplan Yorke dimension}

The Lyapunov exponents are a local property of the system that describes the divergence of nearby trajectories and can give some limited information about the chaotic nature of an ODE system. So by definition given the ode system 
\begin{equation}
\dot{x_i} = f(x_i)
\end{equation}
in a k-dimensional phase space for each coordinate we can define the Lyapunov exponents (spectrum) in the following way:
\begin{equation}
J_{ij}(t) = \frac{df_i(x)}{dx_j}|x(t)
\end{equation}
when \(J_{ij}(t)\) is the Jacobian. It defines the evolution of the tangent vectors given by Y, where we have
\begin{equation}
\dot{Y}=JY    
\end{equation}
with the I.C of \begin{equation}
Y_{ij} = I_0
\end{equation}Y describes the effects of the initial perturbation in x(0) and the end effect on x(t).
We define:
\begin{equation}
\Lambda = lim_{t->\infty} \frac{log(Y(t)Y(t)^T)} {2t}
\end{equation}
the Lyapunov exponents are the respected eigenvalues of lambda.
Now, in practice to calculate the Lyapunov exponents we do the QR decomposition of the matrix that defines the system, with the initial condition 
\begin{equation}
Q_0 = I_0
\end{equation} 
We iterate with time to get the desired Lyapunov exponents, and using the equation for the Kaplan Yorke dimension: 
\begin{equation}
D_{KY} = k + \frac{\sum_i \lambda_i}{|\lambda_{i+1}|}.
\end{equation}
to calculate it. For the calculation, we first order the eigenvalues from the largest to the smallest. \\ The index "k" is the largest index for which the eigenvalue is positive.

\begin{figure}
    \centering
    \includegraphics[width=0.5\linewidth]{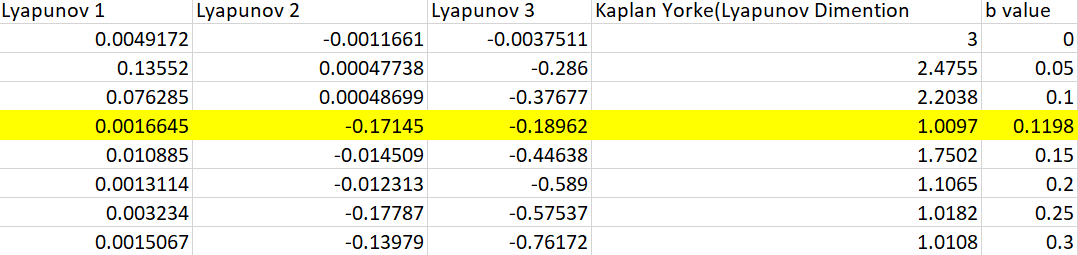}
    \caption{Lyapuov exponents and Kaplan Yorke Dimension for different values of b.}
    \label{fig:enter-label}
\end{figure}

\begin{figure}[htbp]  
    \centering  
    \label{}  
\end{figure}
\begin{figure}
\includegraphics[width=0.5\linewidth]{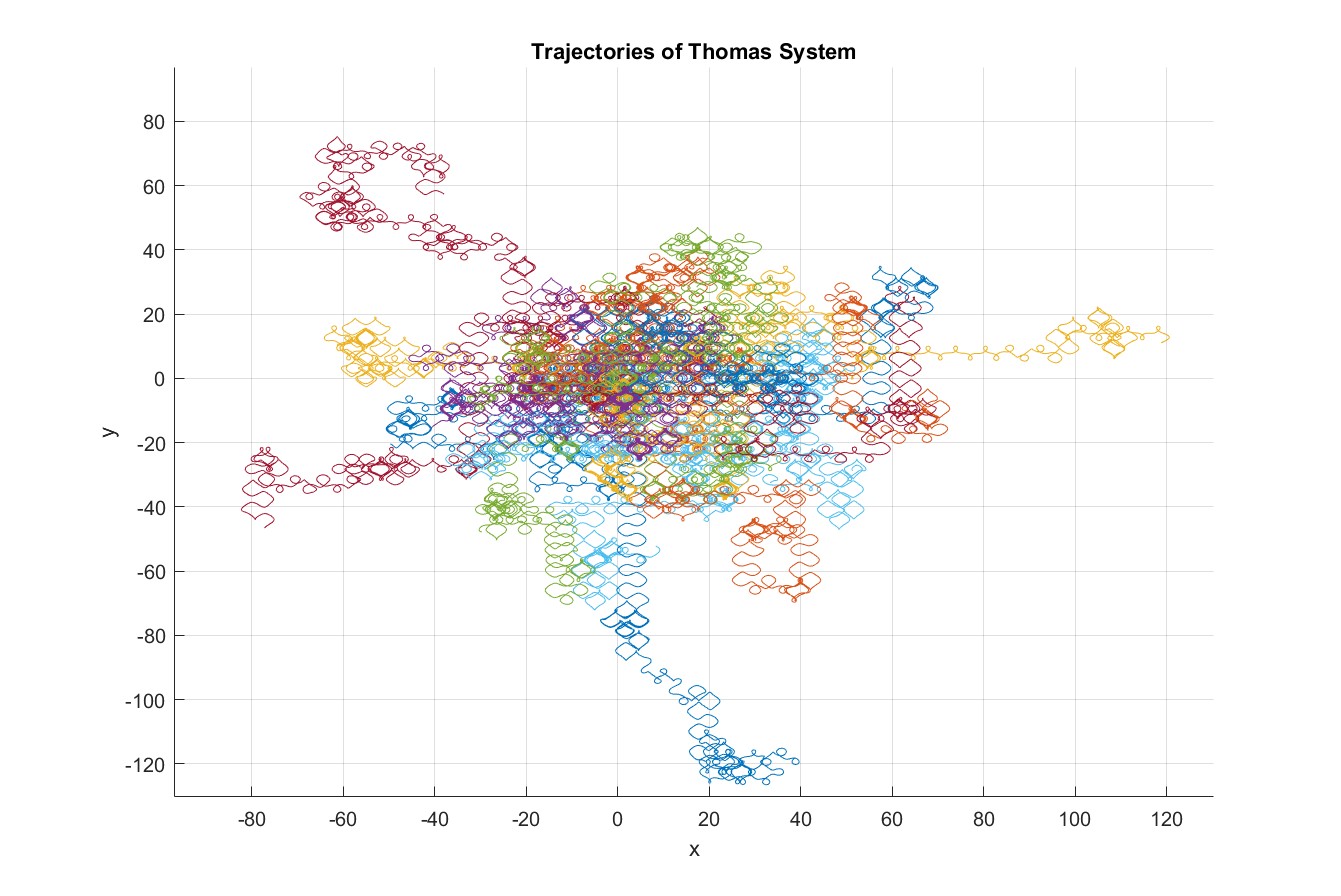}  
\includegraphics[width=0.5\linewidth]{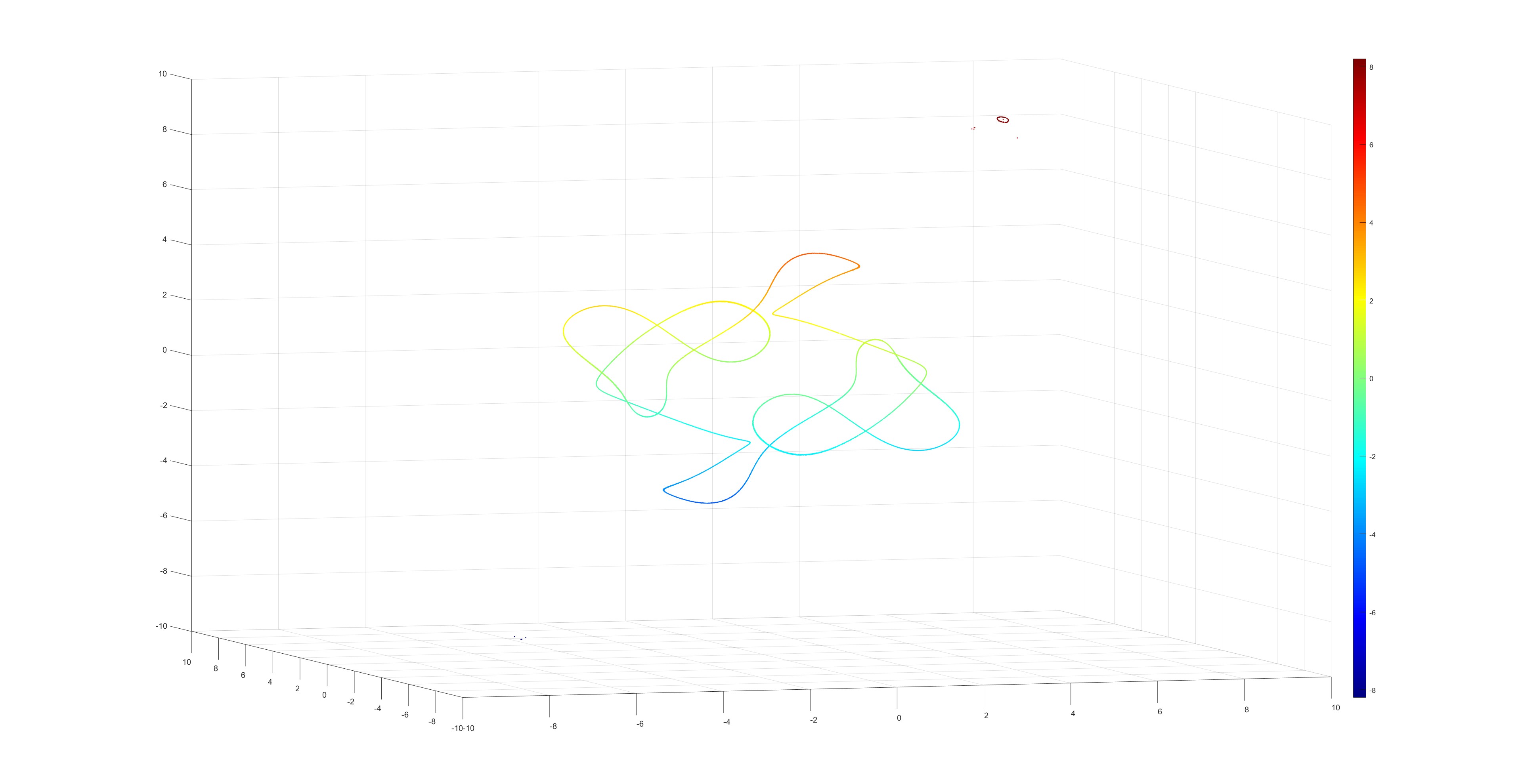}
    \caption{The case of b=0 and of the limit cycle at b = 0.1198 }
    \label{fig:enter-label}
\end{figure}

\begin{figure}[htbp]  
    \centering  
    \label{}  
\end{figure}
\begin{figure}
\includegraphics[width=0.5\linewidth]{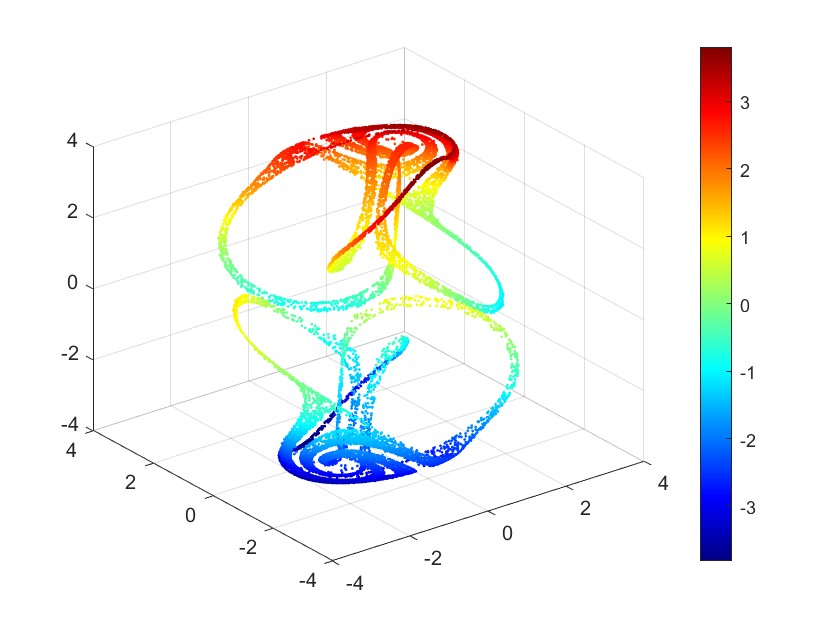}
\includegraphics[width=0.5\linewidth]{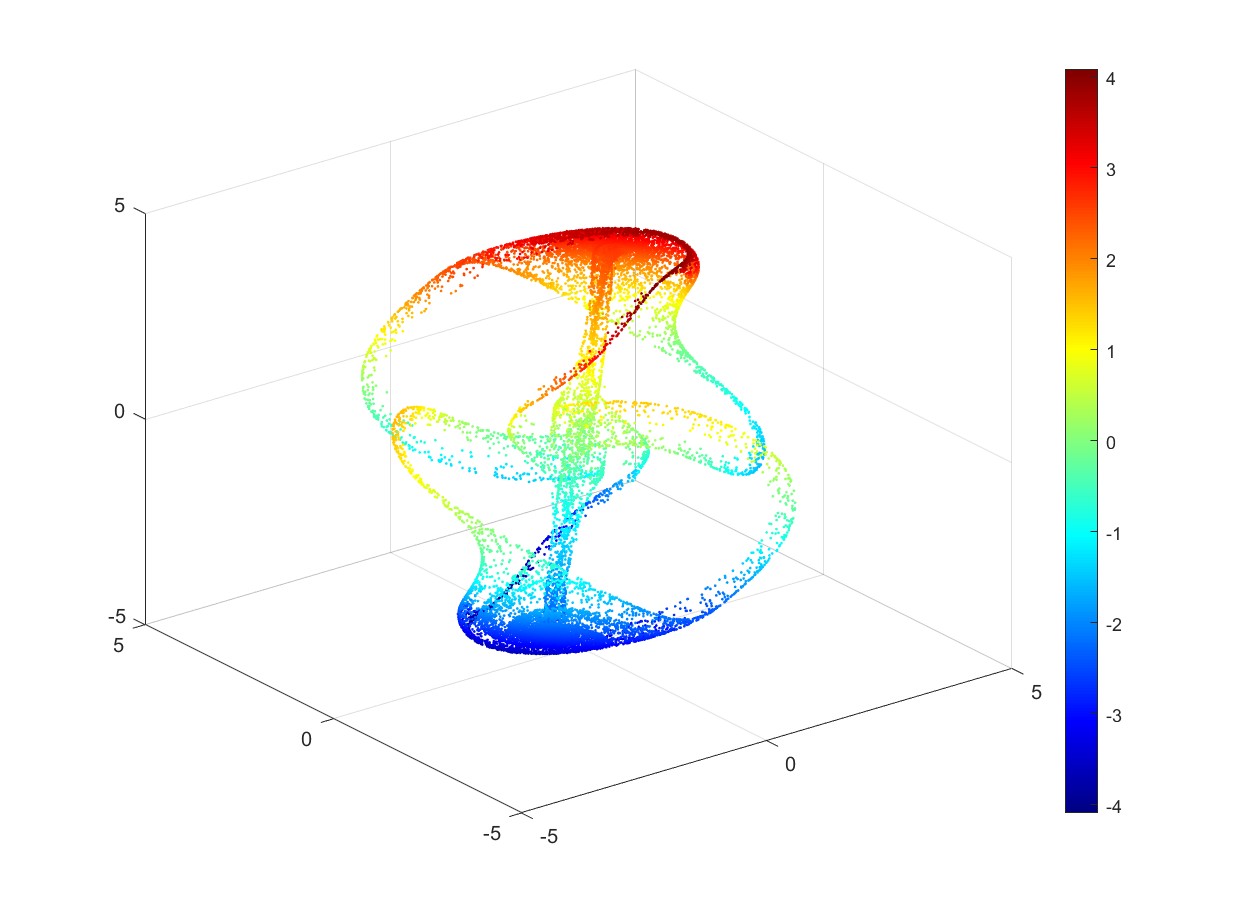}  
\includegraphics[width=0.5\linewidth]{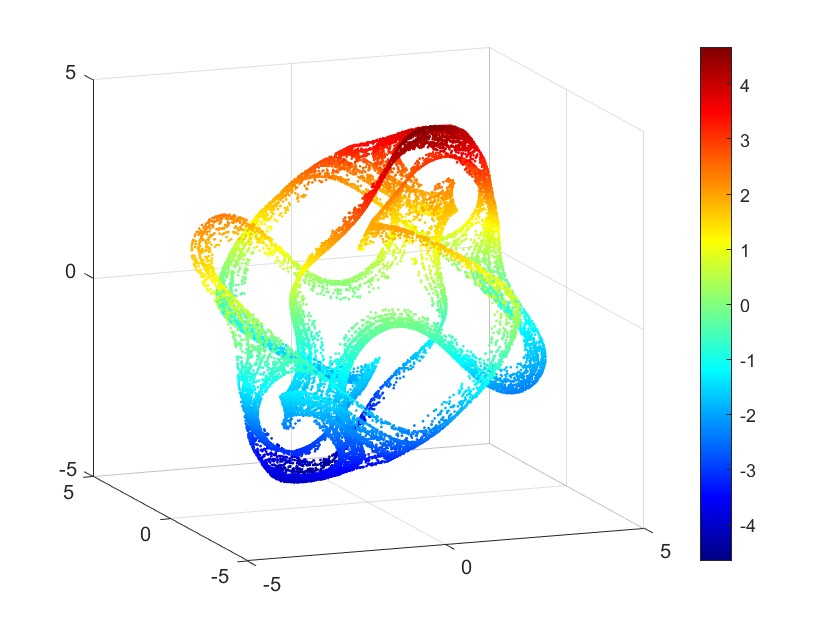}
\includegraphics[width=0.5\linewidth]{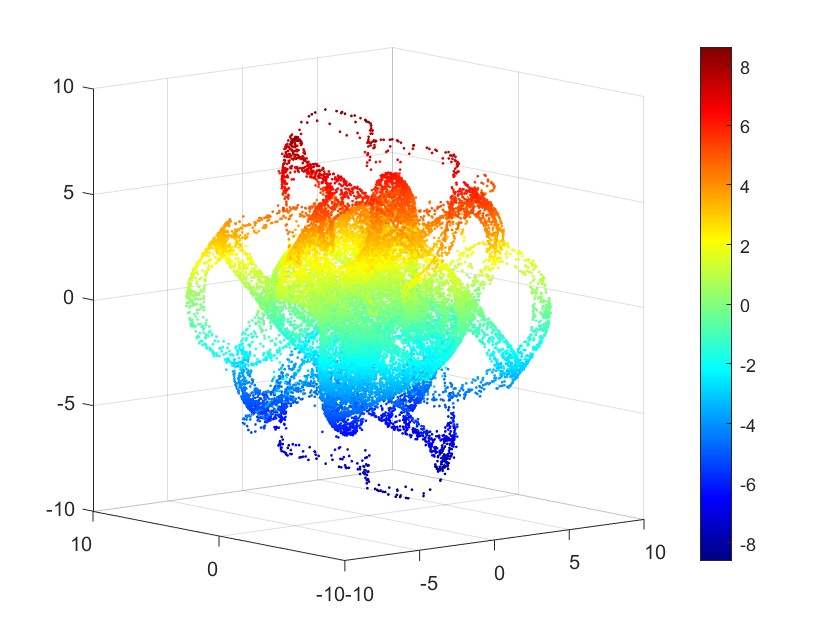}
\includegraphics[width=0.5\linewidth]{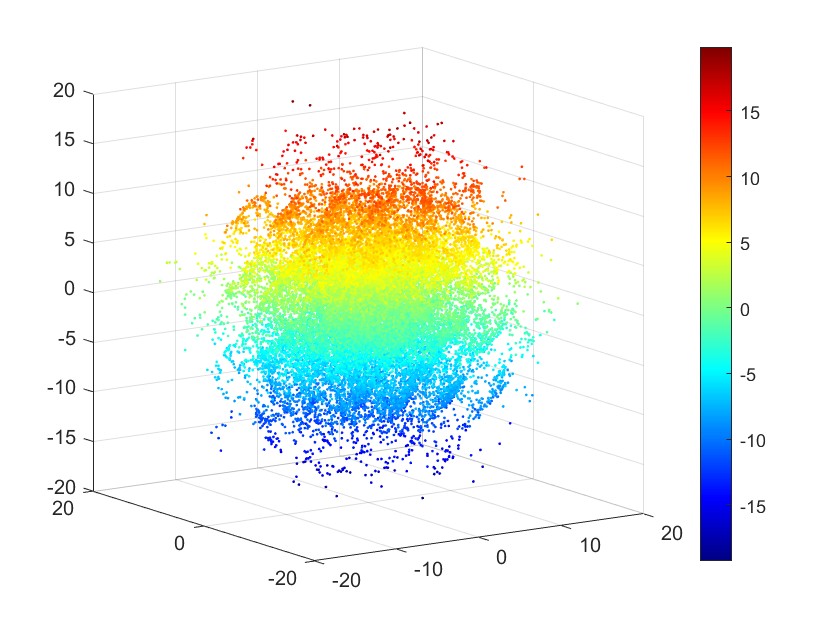}
\includegraphics[width=0.5\linewidth]{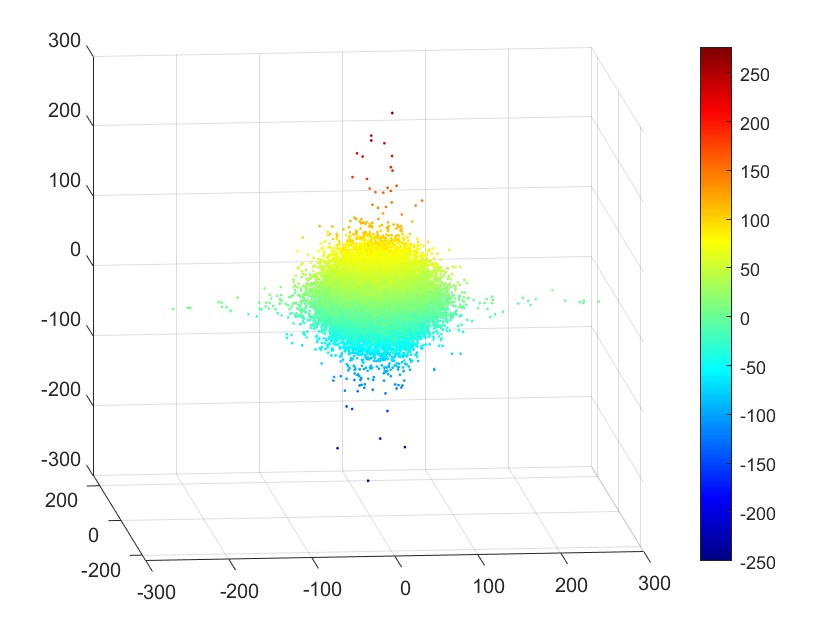}
    \caption{color maps for b = 0.22,0.19,0.15,0.08,0.03,0 and the descendence to chaos.}
    \label{fig:enter-label}
\end{figure}

From Figure 3 and Figure 5, we can see a qualitative change in the behavior of the system, namely that the attractor had split up into two mirror attractors that do not mix with one another. From figure 4 we can see that the dimension of the system rises as b gets smaller but not whenever we have a limit cycle. For example, at b = 0.1198 we see that the dimension drops to around 1, which is logical since for this value of b  there is a limit cycle which is a one-dimensional curve in the phase space. \\
For b larger than 0.33 the dimension of the system approaches the value of 1, since in this region the system is yet to undergo the periodic cascade and is yet to be chaotic. For such b the system converges into two stable points/lines. On the other side of the spectrum, as b approaches zero, the dimension of the system approaches 3, and for b = 0, the system experiences strange behavior as indicated by Figure 3.

\subsection{Bifurcation diagram and the Poincare section}
A Poincare section is a way of detecting limit cycles in an ode system. We construct a plane of interest and search for D-1 periodic behavior on the plane which indicates a limit cycle in the original system. We have here two approaches where each one gives us different information. In Figure 4 we see the Poincare section taken as:
\begin{equation}
bz = sin(x) 
\end{equation} 

We simulate a sample of initial conditions normally distributed in phase space and search for first and second intersections with the function, but as this is not "simple as a plane" it is difficult in some sense to conclude it, although on closer inspection one can deduce some of the properties of the system form the diagram such as symmetry to clockwise permutations as well as symmetry to reflection. On the other hand, the bifurcation diagram shows the intersections again with the plane \(\dot{z} = 0\) but here we simulate for different values of b and we can remarkably well watch as the system approaches chaos. The diagram agrees with our results on the location of different bifurcations which happen for particular values of b which can be seen in the diagram.
\begin{figure}[H]  
    \centering  
    \label{}  
    \includegraphics[width=0.75\linewidth]{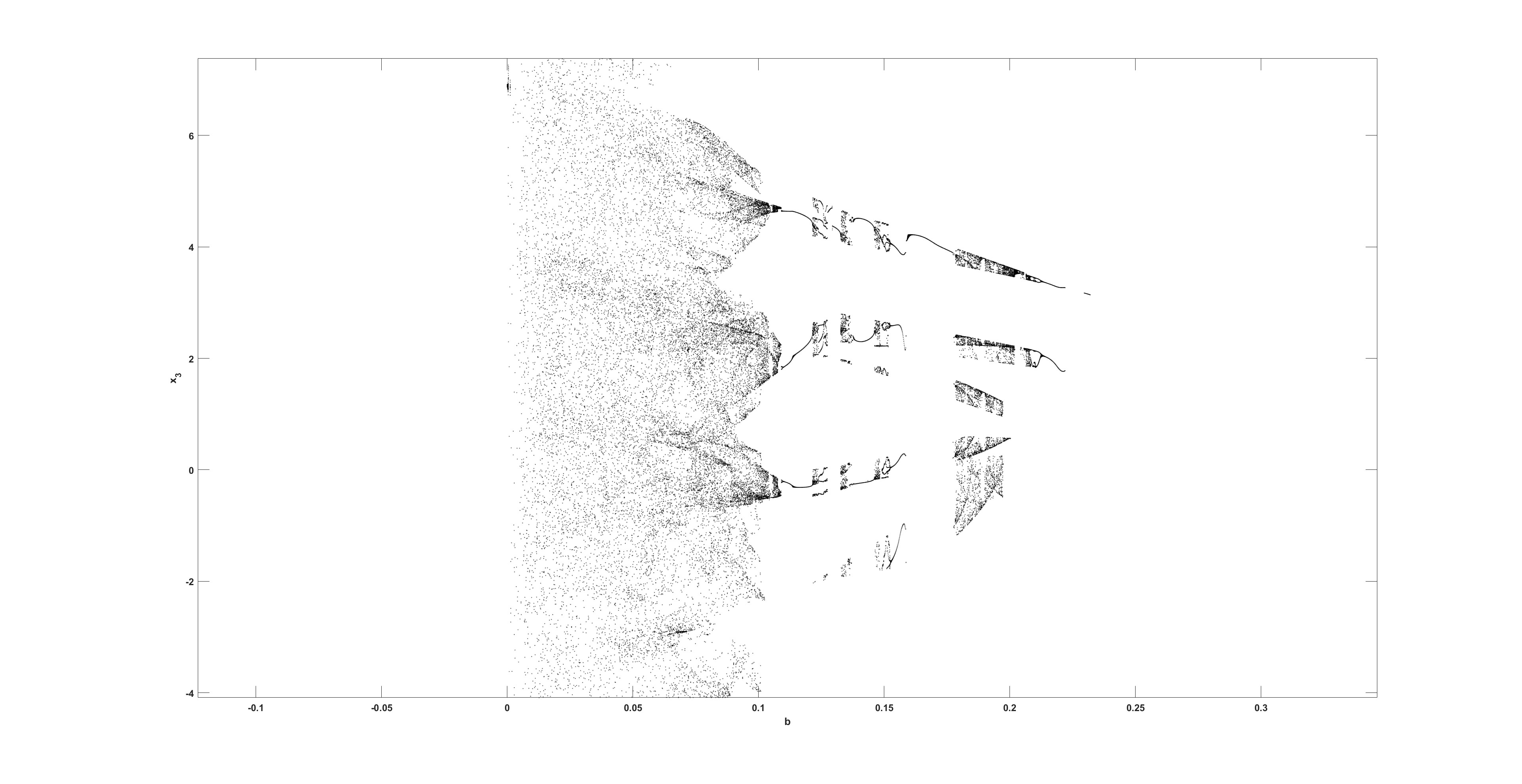}
\end{figure}
\begin{figure}[H]
    \centering  
    \label{}  
    \includegraphics[width=0.55\linewidth]{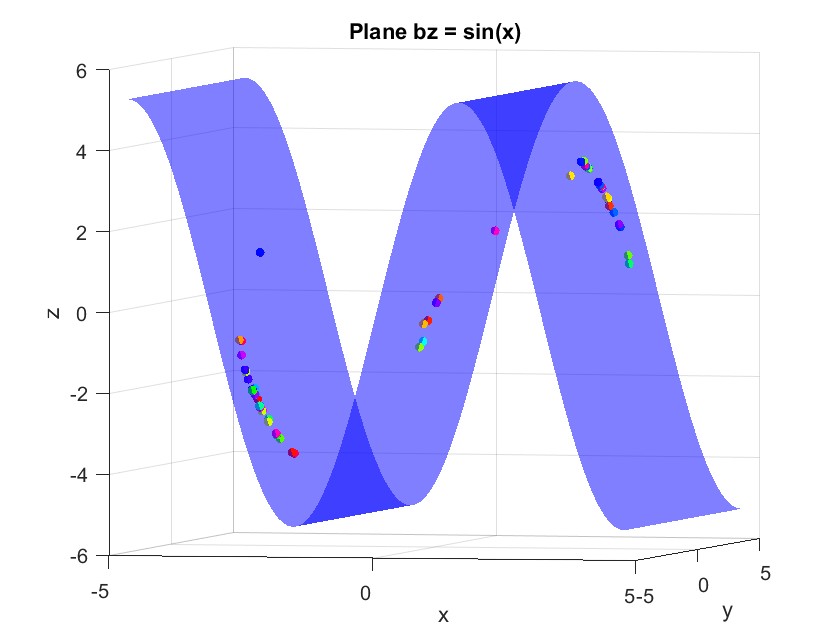}
    \caption{Bifurcation diagram, the intersections with the plane \(\dot{z}\)= 0, Poincare section of bz = sin(x) for b = 0.19}
\end{figure}

\newpage
\section{The case b=0}
In the case of b=0, there is no dissipation in the system, and the system is acting like a Brownian motion Sprott [2] indeed, in the simulation of the system for b=0 it looks like the points are distributed randomly. We can see this special behavior in figures (3,5), where we can see that the system behaves differently in this case. We will now analyze it analytically. \\
The system has now infinity fixed points from the form \[x*=\pm \pi n , \ y*=\pm \pi m \ ,z*=\pm \pi k\] when n,m,k are integers.
The jacobian matrix is 
\begin{equation}
J = \begin{bmatrix}
0 & cos(y) & 0 \\
0 & 0 & cos(z) \\
cos(x) & 0 & 0 \\
\end{bmatrix}
\end{equation}
so for the fixed points the jacobian matrix is:
\begin{equation}
J = \begin{bmatrix}
0 & \pm{1} & 0 \\
0 & 0 & \pm{1} \\
\pm{1} & 0 & 0 \\
\end{bmatrix}
\end{equation}
when the sign is depending on the parity of the integers n,m,k.
Anyway, there are only two possible characteristic polynomials: \[-\lambda^3 \pm{1}\] when taking the positive sign, the eigenvalues are \[ \lambda_0=1 , \lambda_{1,2}=-\frac{1}{2}\pm{\frac{\sqrt{3}}{2}i}\] and when we take the negative sign, the signs of the eigenvalues are reversed, so we get that every fixed point is unstable.
Note that since b=0, \[
\nabla \cdot \textbf{v}  = -3b = 0 
\]
so the density is constant.
Physically speaking, we can also calculate the average speed of the particle:
\[\Bar{v}=\sqrt{\Bar{\dot{x}}^2+\Bar{\dot{y}}^2+\Bar{\dot{z}}^2}=\sqrt{\Bar{sin(x)}^2+
\Bar{sin(y)^2}+\Bar{sin(z)^2}}=\sqrt{\frac{3}{2}}\approx1.2247\] since every term in the square root is: \[\frac{1}{\pi}\int_{0}^{\pi} \sin^2 t \, dt = \frac{1}{2}
\]

\section{Conclusion}

In this paper, we explored a 3-D non-linear dynamical system controlled by a single parameter \(b > 0\). We saw that even though there is only one parameter, rich dynamics is hidden inside the system, including infinite bifurcations, limit cycles, and multiple attractors. We also considered the case of \(b=0\), which is interesting in itself since the system is transformed into Brownian motion and ergodic theory.

\newpage

\section*{References}

\begin{enumerate}
\item Deterministic chaos seen in terms
of feedback circuits: Analysis,
Synthesis, " Labyrinth Chaos"
R. Thomas
Laboratoire de G´en´etique des Procaryotes and
Center for Nonlinear Phenomena and Complex Systems,
Universit´e de Bruxelles, Bruxelles
Received February 25, 1999; Revised April 1, 1999.
    \item J.C. Sprott, University of Wisconsin, Konstantinos E. Chlouberakis, University of Athens, 2006.
    \item Otto E. Rossler, Continuous Chaos-Four Prototype Equations, Institute for Physical and Theoretical Chemistry, University of Tübingen, Institute for Theoretical Physics, University of Stuttgart, 7400 Tübingen, Federal Republic of Germany.
    \item Deterministic nonperiodic flow
E. Lorenz
Published 1 March 1963
Physics, Environmental Science
Journal of the Atmospheric Sciences.
\item Kaplan, J. \& Yorke, J. [1979] “Chaotic behavior of
multidimensional difference equations,” in Functional
Differential Equations and Approximations of Fixed
Points, Lecture Notes in Mathematics, Vol. 730, eds.
Peitgen, H.-O. \& Walther, H.-O. (Springer, Berlin),
pp. 228–237.
E. Lorenz
Published 1 March 1963
Physics, Environmental Science
Journal of the Atmospheric Sciences.
\item Mandelbrot, B. B. [1983] The Fractal Geometry of
Nature (Freeman, San Francisco).
Peak, D. \& Frame, M. [1994] Chaos under Control: The
Art and Science of Complexity (Freeman, NY).
\item Rasmussen, S., Knudsen, C., Feldberg, R. \& Hindsholm,
M. [1990] “The coreworld: Emergence and evolution of
cooperative structures in a computational chemistry,”
Physica D 42, 111–134.
\item Kauffman, S. [1993] The Origins of Order (Oxford University Press).
\item Sprott, J. C. \& Rowlands, G. [2001] “Improved correlation dimension calculation,” Int. J. Bifurcation and
Chaos 11, 1861–1880.
\item Sprott, J. C. [2003] Chaos and Time-Series Analysis
(Oxford University Press), pp. 196–198.
\item Thomas, R., Basios, V., Eiswirth, M., Kruel, T. \&
R¨ossler, O. E. [2005] “Hyperchaos of arbitrary order
generated by a single feedback circuit, and the emergence of chaotic walks,” Chaos 14, 669–674.
\item Wolf, A., Swift, J. B., Swinney, H. L. \& Vastano, J. A.
[1985] “Determining Lyapunov exponents from a time
series,” Physica D 16, 285–317.
\item Tsonis, A. A. [1992] Chaos: From Theory to Applications
(Plenum, NY).
\item Nonlinear dynamics and chaos with applications to Physics, Biology,Chemistry, and Engineering Steven H. Strogatz Published 2018 by CRC Press
Taylor \& Francis Group
6000 Broken Sound Parkway NW, Suite 300
Boca Raton, FL 33487-2742.
\item Deneubourg and Goss [1989] Self-organized shortcuts in the Argentine ant Volume 76, pages 579–581.
\item May, Robert M. (1976). "Simple mathematical models with very complicated dynamics". Nature. 261 (5560): 459–467.

\end{enumerate}

\end{document}